# An Alternative View: Satisfaction of the Four-variable Bell Inequality Using Quantum Correlations

**Louis Sica[1,2]**

[1]**Institute for Quantum Studies, Chapman University, Orange, CA & Burtonsville, MD, 20866, USA**
[2]**Inspire Institute, Inc., Alexandria, VA 22303, USA**

**Email: lousica@jhu.edu**

.

## Abstract

The algebraic derivation of the numerical limits of Bell inequalities in either three or four random variables is independent of the assumption of randomness. The limits of the inequalities follow as mathematical consequences of their created algebraic structures independently of application to random or deterministic variables. The inequalities should be called identity-inequalities. A final correlation re-uses data from the previous correlations and thus leads to the inequality limits. It generally has a different functional form from the previous correlations, whether derived as a counterfactual mathematical result, or in a way enabling comparison with experiment. These algebraic facts and their consequences are central to understanding the inequalities' use, but have not been widely recognized. Logically consistent application of the inequalities to Bell experiments is challenging, given that the number of mathematically assumed random variables is greater than the number of physical variables produced per experimental realization. Given Bell's rejection of the use of sequential, alternative paths, three experimental runs are here considered to enable acquisition of data to be rearranged for computation of statistical cross-correlations. Predicted quantum mechanical correlations then satisfy the inequality. Since mathematically inconsistent use is sufficient to cause inequality violation, the conclusion that violation implies the nonexistence of underlying variables in the entanglement process does not follow.

.

## 1. Introduction

The four variable inequality will be treated here because it is the one most commonly used by experimentalists whose careful diligence has shown that the interactions in the two photon sources used result in cosine correlated counts from any single pair of settings on opposite sides of a Bell experiment [1,2,3,4]. The four variable version of the inequality may be thought to be more applicable to experiments than the three variable inequality because measurements at each of the two setting pairs used in the derivation occur on opposite sides of the apparatus. It was originated by Clauser, Horn, Shimony, and Holt,   (CHSH) [1], and discussed at length by Bell [5].   It is also considerably more complex to analyze in application than the three variable case treated in [6]. Unfortunately, while the Bell inequality and its experimental violation are widely recognized, the true reason for this violation is only understood by a small but growing number of workers.   The basic fact is that Bell inequalities in either three or four variables share an in-obvious logical characteristic: they are founded on algebraic identity-inequalities that cannot be violated by either three or four data sets, respectively, of $\pm 1's$.    Only the form of the correlations among the variables may vary for random or deterministic variables but not their satisfaction of the appropriate inequality if used with logical consistency under the imposed constraints. The reason for this (surprising to many) claim, is that in each case a final correlation that appears to be independently determined in the final expression of the derivation, results from the products of the data-pairs that produced the previous correlations [7]-{11}. This results in correlation functions having new forms except in the special case of spatial stationarity.

In experimental application of the inequality, some of the correlated variables must be obtained in independent runs. They become correlated through a correlation to other correlated variables by what is defined as conditional independence [12] in probability theory. The fact that the variables in question are restricted to ±1 values leads to these overlooked relations. The unexpected correlations that result from conditional independence in the four-variable inequality are a major result of this paper and are analyzed below.

The four-variable inequality will first be derived as a statistics result using Bell's notation assuming random hidden variables and their associated probability density. The algebraic construction using four cross-correlations leads to limits of ±2. The result will then be twice re-derived: first based on a joint probability density without hidden variables, and second, from the assumption of non-random data sets. This will show surprisingly, that the inequality cannot be violated by experimental data used consistently with its structure based on cross-correlation. A further problem then immerges along with a solution: how to apply the four variable inequality to Bell experiments producing only two random variables per realization.

## 2. Inequality Derivation With and Without Bell Hidden Variables

Bell, following CHSH, defined variables $A(a,\lambda), A'(a',\lambda), B(b,\lambda),$ and $B'(b',\lambda)$, each equal to ±1, where $a$, $a'$, $b$, and $b'$, designate instrument setting angles in a Bell-experiment apparatus (see Fig. 1). Labels $A$, and $a$, indicate a readout and corresponding angular setting, respectively, on one side of the apparatus while labels $B$ and $b$ indicate a readout and angular setting on the other side. If $a = b$, $A = -B$ to make the mathematical construction consistent with requirements of entanglement. The $\lambda$ symbol designates random variables that are the same for all instrument settings for each two-photon event, and produces the Bell inequality as a statistical result.

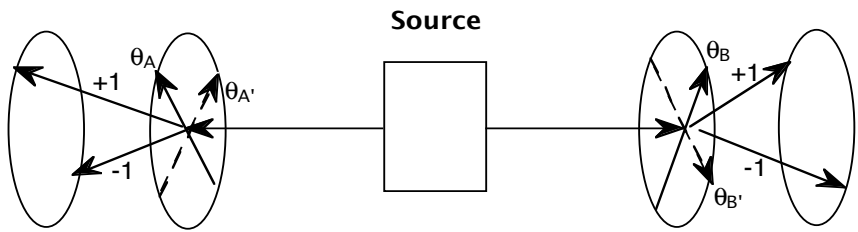

**Fig. 1.** Schematic of Bell experiment in which a source sends two particles (photons most often used) to two detectors having angular settings $\theta_{A'}$ and $\theta_{B'}$, (denoted as $a$ and $b$ in Bell's notation) and alternative settings $\theta_{A'}$ and $\theta_{B'}$. While one measurement operation on the $A$-side, e.g. at setting $\theta_{A}$, commutes with one on the $B$-side at $\theta_{B}$, any additional measurements at either $\theta_{A'}$ or $\theta_{B'}$ are non-commutative with prior measurements at $\theta_{A}$ and $\theta_{B}$, respectively. The figure, drawn by the author. has been modified for use in various papers.

The purpose of Bell's construction followed by its violation (due to mathematically inconsistent use) was to investigate the implication of assuming hidden variables $\lambda$ unknown to quantum mechanics to account for predicted entanglement-based statistical correlations. Correlations among multiple measurements at alternative instrument settings on two particles were assumed. The four-variable Bell inequality originated by CHSH [1] [5] is:

$$-2 \le \int A(a,\lambda)B(b,\lambda)\rho(\lambda)d\lambda + \int A(a,\lambda)B(b',\lambda)\rho(\lambda)d\lambda +$$
$$\int A(a',\lambda)B(b,\lambda)\rho(\lambda)d\lambda - \int A(a',\lambda)B(b',\lambda)\rho(\lambda)d\lambda \le 2 \,, \tag{2.1}$$

with the explanation of its limits equal to $\pm 2$ to follow. Given that the values of $\lambda$ are the same for each pair of functions in the four integrals, (2.1) may be written

$$-2 \le \int \Big( A(a,\lambda)\big[ B(b,\lambda)+B(b',\lambda)\big] + A(a',\lambda)\big[ B(b,\lambda)-B(b',\lambda)\big] \Big)\rho(\lambda)d\lambda \le 2 \,.\` \tag{2.2}$$

The integrand factor in curved parentheses equals + or − 2 for each $\lambda$ -value since one expression in rectangular brackets must equal $\pm 2$ and the other zero, depending on whether $B(b,\lambda_i)$ and $B(b',\lambda_i)$ have the same or opposite signs at a given value $\lambda_i$, Thus,

$$-2 \le A(a,\lambda_i)\big[ B(b,\lambda_i)+B(b',\lambda_i)\big] + A(a',\lambda_i)\big[ B(b,\lambda_i)-B(b',\lambda_i)\big] \le 2 \,. \tag{2.3a}$$

The integral (2.2) would be maximized or minimized if the integrand factor in (2.3a) remained a constant of either plus or minus 2 for all $\lambda$. The value of (2.2) would then depend only on the probability integral equal to 1:

$$-2 \le \Big( A(a,\lambda_i)\big[ B(b,\lambda_i)+B(b',\lambda_i)\big] + A(a',\lambda_i)\big[ B(b,\lambda_i)-B(b',\lambda_i)\big]\Big)\Big( \int \rho(\lambda)d\lambda \Big) \le 2 \tag{2.3b}$$

However, the same limits are obtained if the random variable readouts are predicted by a joint probability density $\rho(A,B,A',B')$, yielding the integral

$$\int \Big( A(a)\big[ B(b)+B(b')\big] + A(a')\big[ B(b)-B(b')\big]\Big)\rho(A,B,A'.B')dAdBdB'dA' \tag{.2.3c}$$

This has the same numerical limits as (2.1) without the assumption of hidden variables $\lambda$. The form of $\rho(A,B,A',B')$ depends on the experimental procedure used to obtain the four variables in the integrand. Since, in the quantum mechanical case, the alternative measurements at $b$ and $b'$, and $a$ and $a'$, do not commute and cannot physically exist simultaneously in the experiments under consideration, a special procedure for obtaining data to which (2.3c) is applicable is developed below. Neither relation (2.1) or (2.3c) generally applies to measurements obtained in four independent realizations as has been used in practice.

A crucial, almost obvious, but little known result further emphasizes the purely algebraic nature of the inequality independently of both the assumptions of hidden variables and randomness. Consider the inequality

$$-2 \le a_i(b_i+b'_i)+a'_i(b_i-b'_i) \le 2 \,,\` \tag{2.3d}$$

where the subscripted variables are defined to have specific values of $\pm 1$ corresponding to different variables' instrument settings $a, a', b, b'$. Upon inspection, it is seen to hold for both deterministic variables and random variables. It emerges that the basic relation underlying the Bell inequality is an identity-inequality that must be satisfied by any four variables, random, deterministic, local, or nonlocal, i.e., variables that may even be functions of each other (as long as they are employed algebraically as in (2.3d)). The variables can be generated from unrelated physical or mathematical processes as well as made up nonsense, and (2.3d) will be satisfied as long as the four variables used each equal $\pm 1$. If more than four variables are used, ignoring the basic algebraic structure, the inequality may be violated, just as a trig-function identity in the difference of two angles may be violated if six angles are inserted into the equality.

What makes these facts even more central to critical examination of the Bell theorem is that in experiments, the correlations in (2.1) are not physically measured or observable, only finite sets of random counts are observed, and compared with predicted results to which they are assumed to statistically converge. From a number $N$ of counts for each variable in (2.3d) one obtains

$$-2N \leq \sum_i^N a_i(b_i + b'_i) + a'_i(b_i - b'_i) \leq 2N \quad , \tag{2.3e}$$

or after dividing by $N$:

$$-2 \leq \frac{1}{N} \sum_i^N [a_i(b_i + b'_i) + a'_i(b_i - b'_i)] \leq 2 \quad . \tag{2.4}$$

For very large numbers of counts [5] in the random case, the average correlations occurring in (2.4) are assumed to statistically converge to functions that may be computed from random variable probabilities.

It follows that while (2.1) is derived on the assumption of four variables using random inputs from the same probability density $\rho(\lambda)$, and (2.3c) is derived assuming random variables and probabilities without hidden variables, (2.4) is a purely algebraic result that holds without the assumption that the variables are even random. If they are all $\pm 1's$, then (2.4) must hold. Thus, result (2.4) is the most important of the above inequalities, as it is the basis of the others. One arrives at the conclusion that neither the Bell inequality nor its satisfaction depends logically on the existence of hidden variables. The basis of the Bell theorem is an unrecognized inequality-identity that has not yet been included in standard math tables.

The inequality imposes an algebraic constraint on the correlations leading to its satisfaction. The first three variable pairs occurring in (2.4) determine the fourth variable pair (the third pair is determined by the first two in the three variable case [6]):

$$a_i b_i a_i b'_i a_i'.b_i = a'_i b'_i \tag{2.5}$$

This purely algebraic result has been noted previously by Redhead [7], but without appreciating its significance. Hess found it previously discovered by Vorobov [8,9] in the context of random variables. It has been extensively discussed by Lad [10] who pointed out that four such relations can be obtained from the inequality since each one can be generated from the other three. (Depending only on basic algebra, consequences following from this were given in [13]). Thus, while one may determine the first three correlations in the four-variable Bell inequality by choosing to measure the specified pairs in a consistent manner to be shown, the fourth correlation is determined from the data pairs in (2.5), and leads to satisfaction of the inequality with limits ±2. In general from (2.5), one would expect that if the first three correlations originated from one source of interaction, the fourth correlation would in general have a different functional form since it is created by a different mechanism. In the random case to be shown, the variables are conditionally correlated to each other due to their correlations with variables previously observed. Only in the special case of spatial stationarity would the constraint (2.5) result in a correlation having the same form as the previous correlations. However, spatial stationarity is inconsistent with the non-commutation that applies in the Bell case.

From the above analysis of the algebraic basis for the Bell inequality in the four variable CHSH form, it should be clear that if statistically predicted correlations of laboratory data violate it, an error is implied. Given that for any four finite data sets, (2.4) must be identically satisfied, no predicted correlations that violate the inequality can correspond to four actually existing data sets. Further, it cannot matter whether the data are measured or mathematically created, violation of (2.4) would imply violation of basic principles of mathematics. It must be noted that it is (2.4) that is directly related to physical experiments and not the theo-retical counterparts (2.2) and (2.3c). How the necessary correlations may be computed to enable application of the Bell inequality is described next. It is an extended version of that given for the three-variable case [14].

## 3. How Can a Four Variable Inequality Be Applied to a Random Process Yielding Only Two Outputs Per Realization?

### 3.1 Logically Consistent Bell Counterfactuals Must Still Satisfy the Bell Inequality

The Bell inequality, assuming the Bell hidden variable notation of (2.2), cannot be applied to experimental results. This follows from the fact that the same hidden variable values determine each correlation. If the outputs at a given pair of settings are observed, no observation at alternative settings with the same hidden variables is possible. One cannot undue an experimental observation at a given setting and repeat the observation at a different setting assuming that the random processes leading to the first readout repeat with infinite precision.

It has sometimes been assumed that all correlations contributing to the Bell inequality have the same form because any one of them if observed, has that form. However, (2.5) must still hold and determines that the final pair of outputs leading to inequality limits is identically equal to the product of the first three. Thus, if the first three correlations are mathematically constructed and have the Bell cosine form when averaging over multiple hidden variable sets, the final correlation would be expected to have a different form as it results from products of previously generated variable pairs. Hence, the purely algebraic requirements (2.4, 2.5) underlying the Bell inequality cannot be avoided by using a counterfactual-mathematical interpretation based on hidden variables. These considerations imply that although any one observed correlation has a Bell cosine form, the four taken together must have at least one different functional form for the Bell inequality to be applicable.

An alternative mis-interpretation of the physical situation is that all the correlations have the same form because the process is spatially stationary. In that case, (2.5) would lead to the same correlation function as resulted from the previous measured output pairs. Spatial stationarity has been used as a simplifying idealization in some areas of optics and implies that any number of measurement pairs may be obtained in any order since all correlations are the same. However, in the Bell case the measurements involved are non-commutative, with each measurement in a sequence probabilistically conditional on the one before. This has been shown to produce different correlational forms among the variables in the sequence [15]. If assumed to hold in the Bell case, spatial stationarity leads to the insertion of four independent pairs, or eight variables, into a four variable identity-inequality so as to cause its violation. The resulting violation has then been attributed to a construction based on the assumption of hidden variables as in (2.2), followed by the conclusion that therefore such variables cannot exist. This reasoning is now seen to be incorrect.

Note that a computer model calculation of a Bell cosine correlation has been reported in [6, 15] where common information is given to two independent computers. Other examples probably exist in a literature of thousands of papers.

## 3.2. Operational Procedure for Application of the Inequality to an Experiment in the Four Variable Case, and its Satisfaction by Quantum Correlations.

In view of the mathematical facts presented in Sec. 2, it is necessary to reconsider the conditions under which the Bell inequality has been violated in experimental practice. How can a four variable expression shown to be an identity-inequality be violated? It must be assumed that results (2.4) and (2.5) demonstrated above, are known to very few researchers. Since, the inequality is commonly thought to result specifically from the Bell assumptions and notation specifying hidden variables in (2.1), all the correlations whether observed or not, have been assumed to have the same functional form. But as noted, (2.5) must hold even for four mathematically constructed unobserved counterfactual correlations. The final correlation would then have a different functional form from the preceding ones.

The Bell inequalities are algebraic identities constructed from three or four variables, respectively, all present at the time of computation so as to enable calculation of cross-correlations. Given that (2.4) holds as a fact of algebra independently of assumptions of hidden variables, a procedure for its application using quantum correlations consistent with its structure will now be described. It is an extended version of that used in the three variable case [14] and ultimately must be devised to be consistent with (2.4) and (2.5), the basis for all inequality derivations.    Rewriting the variables to be averaged in the Bell inequality in terms of an implied joint probability that must ultimately be specified by quantum mechanics,

$$\sum_{a,b,b',a'} (a_i b_i + a_i b'_i + a'_i b_i - a_i b_i a_i b'_i a'_i b_i) P(a_i, b_i, a'_i, b'_i) = \tag{3.1}$$

$$\sum a_i b_i P(a_i, b_i, a'_i, b'_i) + a_i b'_i P(a_i, b_i, a'_i, b'_i) + a'_i b_i P(a_i, b_i, a'_i, b'_i) - a_i b_i a_i b'_i a'_i b_i) P(a_i, b_i, a'_i, b'_i).$$

The first three terms produce Bell correlations (in the optical case) in the form $-\cos 2(a-b)$ with $a'$ and $b'$ substituted for a and b as appropriate.    For an inequality structure depending on the simultaneous existence of the four variables at the time of evaluation of the inequality, all the variables in probability $P(a,b,a',b')$ must be summed over.    The last term average is

$$\sum abab'a'bP(a,b,a',b') = \sum a'b'P(a',b',a,b) = \tag{3.2}$$

$$\sum a'b'P(a',b',|a,b)P(a,,b) = \sum a'b'P(a'|b)P(b'|a)P(a,,b)$$

after using the fact that the correlation of $a'$ and $b'$ must be conditional on a and b individually (discussed in the Appendix) with which they will have been measured and correlated.    The evaluation of (3.2) after some computation outlined in the Appendix is found to be

$$C(a',b') = -\cos 2(b'-a)\cos 2(a'-b)\cos 2(a-b). \tag{3.3}$$

The procedure for obtaining the data follows from the mathematical form of the Bell inequality and is similar to that used in the three variable case.    First measurements of the pair (a, b) must be obtained in a conventional Bell experiment and arranged in an (infinite) list of pairs.    Theoretically, the probability of observing $\pm 1$ for each variable alone is ½ as will be assumed to hold here.    The same is true for pairs $(a',b)$ and $(b',a)$.    Now under the assumption of ideal measurements (for which all single variables have average zero), experimental data pairs $(a',b)$ can be rearranged so that the b-output values match those in data pair list $(a,b)$.    Similarly, the data pair list for $(b',a)$ can be rearranged so that a-values match those of the list for (a, b).    The positional shifts in the lists are not expected to be large since the probability of occurrence of $\pm 1$ for each variable is the same and equal to 1/2. The result is that three lists of data pairs are now arranged so that the $a$-values and $b$-values in the three paired lists are the same, though measured in three individual (ideal) experiments.    The correlations $C(a,b')$ and $C(b,a')$ may now be computed from their conditional dependence on $C(a,b)$ leading to the result (3.3) as shown in more detail in the Appendix.

## 4.. $C(a',b')$ and the Other Correlations Satisfy the Bell Inequality

The Bell inequality may now be written:

$$-\cos 2\theta_1 - \cos 2\theta_2 - \cos 2\theta_3 + \cos 2\theta_1 \cos 2\theta_2 \cos 2\theta_3 \le 2, \tag{4.1}$$

where the theta values equal differences of polarization beam splitter settings in a Bell experiment.    The cosine terms are each equal to: $\cos 2\theta = 1 - 2\sin^2\theta$.    Replacing $\sin^2\theta$ by x in (4.1) yields

$$-(1-2x_1)-(1-2x_2)-(1-2x_3)-(1-2x_1)(1-2x_2)(1-2x_3) \le 2 \tag{4.2}$$

that after some algebra and rearrangement becomes

$$x_1 x_2 + s_1 x_3 + x_2 x_3 - 2x_1 x_2 x_3 \le 1 . \tag{4.3}$$

Since $0 \leq x_i \leq 1$, (i = 1,2,3) the x's may be replaced with $\in$'s with each $\in$ specified by $x_i = 1 - \in_i$ where $0 \leq \in \leq 1$. Then (4.3) becomes

$$- \in_1 \in_2 \; - \in_1 \in_3 \; - \in_2 \in_3 \; + 2 \in_1 \in_2 \in_3 \leq 0 \; , \qquad (4.4)$$

that in turn may be rewritten

$$-[\in_1 \in_2 \; (1 - \in_3)] - [\in_1 \in_3 \; (1 - \in_2)] - [\in_2 \in_3] \leq 0 \; . \qquad (4.5)$$

Each of the three bracketed terms in (4.5) is positive with a minus sign before it so that the inequality is satisfied.

## 5. Conclusion

The Bell inequalities in three and four variables were created using a hidden variables notation to prove that their satisfaction was incompatible with correlations resulting from entanglement. However, in both the three and four variables cases the final expressions derived are found to be algebraic identity-inequalities that if used consistently with their derivations cannot be violated regardless of whether the variables to which they are applied are random or deterministic. The fact that the inequalities must be identically satisfied if used with mathematical consistency has not been widely recognized, due to the statistical formalism from which they have been derived. However, the inequalities' numerical limits result from algebraic facts that are independent of whether the variables are random or deterministic. The logical reasoning relevant to the possible existence of hidden variables is completely changed as a result of this fact.

In practice, the Bell inequalities have been violated due to mathematical inconsistencies in their use. Eight variables have been inserted into identity-inequalities derived from four variables. Common trig identities could be violated under analogous conditions of use. The underlying misconception responsible is that a pair of measurements at any pair of variable settings in the inequality would yield a Bell correlation, indicating that all correlations are the same regardless of how or whether they are measured. This reasoning is fatally flawed since even a purely mathematical construction of the measurements implies that the fourth correlation depends on products of data pairs obtained in previous constructions so as to generally yield a different correlational form.

A suggested way to make experimental measurements consistent with the structure of the four variable inequality requires three experimental runs to obtain correlations using different setting pairs. Data pairs may then be rearranged so that variables at the same settings have random but equal values as required by the inequality algebra. This is facilitated by the fact that all individual variables have zero mean. Each correlation retains the conditionality of its two variables, and the final correlation in the inequality is determined from previously correlated pairs and the algebra of $\pm 1's$. The result is tedious to compute and is given in the appendix.

The evaluation of the inequalities consistent with their mathematical structure using quantum mechanical correlations results in inequality satisfaction independently of the existence of hidden variables. It must be observed however, that if the Bell theorem is fatally flawed, its converse (that hidden variables exist), does not necessarily follow. Although mathematical counter-examples to the Bell theorem have been given that imply that the mechanism of entanglement is not the sole source of cosine correlations in optics, computational counter examples do not ultimately settle the complex issues in this case that depend on understanding the nature of photons.

## Appendix

Relation (3.3) will now be derived. The derivation depends on a specific method of application of the Bell inequality to be consistent with (2.4). As shown above, the inequality limits in the random variables case result from either a hidden variable representation or a joint probability representation of observables, since in either case the final correlation results from the reuse of already acquired variable pairs with all variables equal to $\pm 1$.

To obtain four random variables for cross-correlation from a physical process producing only two variables per realization requires some special effort. The procedure uses the theoretical assumption that each variable has zero mean, given that the probability that any variable equals +1 or − 1 equals ½. Thus, while the variables occur in correlated pairs, they are individually completely random. That implies that if two pairs are acquired with a setting in common, for example $(a,b)$ and $(a,b')$, the $(a,b')$ pair order may be rearranged so that the order of $\pm 1's$ of the $a$-values is the same as that in the list for $(a,b)$ while keeping the companion $b'$ value attached. The correlation $C(b,b'|a)$ for $a = +1$ and $a = -1$ may now be computed. The set of pairs for $(b,a')$ may similarly be rearranged so that the order of $b$-outcomes (with each accompanying $a'$ -value remaining attached) is the same as that of list $(a,b)$. The probabilities characterizing the data sets are applicable and enable calculation of the probability $P(a',b')$ :

$$P(a',b') = P(a',b'|a,b)P(a,b) = P(a'|b)P(b'|a)P(a,b) \quad . \tag{A.1}$$

Thus though $a'$ and $b'$ have been acquired in separate experimental realizations, they are correlated because $a'$ is correlated to $b$, and $b'$ is correlated to $a$, with $a$ and $b$ correlated to each other. It should be emphasized that the factorization of $P(a',b'|a,b)$ used in (A.1) is an assumption that follows from the procedure used to acquire and organize the data to enable application of the inequality, plus the physical condition of entanglement. No other physical influences are assumed present. The four variables necessary for application of (2.5) are now available for cross correlation, four-variable data-set by four-variable data-set. This data-set system now satisfies the conditions for applicability of (2.4) and (2.5) allowing the computation of $C(a',b')$.

The computation will now be outlined. In terms of the probability (A.1) the correlation $C(a',b')$ is

$$C(a',b') = \sum_{a_i,b_i} \left[ \; a_+'b_+'P(a_+'|b_i)P(b_+'|a_i)P(a_i,b_i) + a'_-b'_- \, P(a_-'|b_i)P(b_-'|a_i)P(a_i,b_i) \right.$$
$$\left. + a_+'b_-'P(a_+'|b_i)P(b_-'|a_i)P(a_i,b_i) + a_-'b_+'P(a_-'|b_i)P(b_+'|a_i)P(a_i,b_i) \; \right] . \tag{A.2}$$

The contributions corresponding to each sign combination of $(a',b')$ will be numbered (1), (2), (3), and (4). Terms (1) and (2) are equal with (1) given by

$$(1) = \frac{1}{2}\sin^2(a-b)\left[ \sin^2(a'-b)\sin^2(b'-a) + \cos^2(a'-b)\cos^2(b'-a) \; \right] +$$
$$\frac{1}{2}\cos^2(a'-b)\left[ \cos^2(a'-b)\sin^2(b'-a) + \sin^2(a'-b)\cos^2(b'-a) \; \right]. \tag{A.3}$$

Term (3) is given by

$$(3) = -\frac{1}{2}\sin^2(a-b)\left[ \sin^2(a'-b)\cos^2(b'-a) + \cos^2(a'-b)\sin^2(b'-a) \; \right]$$
$$-\frac{1}{2}\cos^2(a-b)\left[ \cos^2(a'-b)\cos^2(b'-a) + \sin^2(a'-b)\sin^2(b'-a) \; \right], \tag{A.4}$$

with (4) equal to (3). The sum of terms (1) through (4) equals $C(a',b')$ or (3.3).

## Acknowledgements

The presentation of the material given above has been influenced by many discussions of the issues with Joe Foremen, personal communications with Karl Hess and Armen Gulian, and critiques of earlier papers by Michael Hall and Frank Lad.